\documentclass[12pt]{article}
\begin{document}
\title{Why the astrophysical black hole candidates are not rotating black holes}
\author{Abhas Mitra\\
Nuclear Research Laboratory, Bhabha Atomic Research Centre,\\
Mumbai -400085, India\\
Email: amitra@apsara.barc.ernet.in, abhasmitra@rediffmail.com}
\date{}
\maketitle
It is  believed that the basic component of the central engine of quasars,
micro-quasars, and energetic  Gamma Ray Bursts  are the rotating or the
Kerr Black Holes (BH)[1]. But by using a generic property[2-4] of the
metric components of a stationary axisymmertic rotating metric in its {\em
standard form}, namely, $g_{\phi \phi} =
\sin^2 \theta g_{\theta\theta}$, where $\phi$
is the azimuth angle and $\theta$ is the polar angle measured from the
axis of symmetry, we have found the unexpected and surprising result that
(i) in order to have a mass of a Kerr BH $m \ge 0$, it is necessary that
its rotation parameter  $a=0$ and if one insists for an $a \ge 0$, one
must have $m\le 0$! Thus if the suspected Black Hole candidates with $m
>0$ are really rotating they cannot be BHs at all which is in agreement
with some detailed analysis of recent observations[5-8].  However, if it
is assumed that such objects are strictly non-rotating, they could be
non-rotating Schwarzschild BHs ($a=0$) with $m\ge 0$ if we ignore the
physical difficulties associated with the existence of such objects.  This
result  calls for new theoretical efforts to understand a  vast range of
astrophysical phenomenon. If one derives the Kerr Metric in a
straightforward manner by using the Backlund transformation, it is seen
that $a= m \sin \phi$. This relationship confirms that $a=m=0$ for Kerr
BHs.

To describe  the axially symmetric stationary 
Kerr spacetime or any other axially symmetric 
stationary spacetime, it is convenient
to consider $x^0 =t$, the time coordinate, and $x^1= \phi$, the azimuth
angle. By definition, for such a spacetime,
the metric coefficients are independent of $t$ and $\phi$:
\begin{equation}
g_{ik} = g_{ik} (x^2, x^3)
\end{equation}

Further for a  spacetime rotating with increasing $\phi$, it is also
required that the spacetime is invariant under simultaneous inversion of of
$t$ and $\phi$, i.e, under the transformation $t \rightarrow -t$ and
$\phi \rightarrow -\phi$[2-4]. This demands that
\begin{equation}
g_{t2} =g_{t3} = g_{\phi 2} =g_{\phi 3} =0
\end{equation}
This brings the metric in the following form:
\begin{equation}
ds^2 = g_{tt} dt^2 + g_{\phi \phi} d\phi^2
+ g_{22} (dx^2)^2   + g_{33} (dx^3)^2 + 2 g_{\phi t}
d\phi dt + 2 g_{23} dx^2 dx^3
\end{equation}
Further, this form of a metric remains unchanged under a coordinate transformation
of the form[2-4]
\begin{equation}
x^2 = x^2 (x^{2\prime}, x^{3\prime}), \qquad x^3 = x^3 (x^{2\prime}, x^{3\prime})
\end{equation}
The above constraints, in turn, imply additional constraints on the metric, and
in particular,
 if we choose spherical polar coordinates with $x^2= r$ and $x^3 =\theta$,
then we should have ({\em see  Eqs. (3.3) and (3.30) in Ref.[2] and  Eq. (74)
 in Ref. [3]})
\begin{equation}
g_{23} = g_{r \theta} =0
\end{equation}
and
\begin{equation}
g_{\phi \phi} = g_{\theta \theta} \sin^2 \theta
\end{equation}
so that eventually the metric is of the form
\begin{equation}
ds^2 =  g_{\phi \phi} d\phi^2 + g_{rr} dr^2 
 + g_{\theta \theta} d\theta^2 + 2 g_{\phi t} d\phi dt +g_{tt} dt^2 
\end{equation}
It can be seen that Eq.(6) incorporates the physical 
condition of asymptotic freedom of the metric[2] 

``that the star and its manifold are axially symmetric; i.e. that 
there exists a Killing-vector field, ${\bf \xi}_\phi$, with closed
orbits, which, at radial infinity, is spacelike and is orthogonal to
${\bf \xi}_t$ and has length $r \sin \theta$.''

However, note that while $g_{\phi \phi}$, $g_{tt}$ and $g_{\phi t}$ are
invariant under arbitray tansformation of the form (4), $g_{\theta \theta}$
in Eq.(6) corresponds to a unique choice of $\theta$[2]
 Suppose  we subject the metric(7)
to the following coordinate transformation (S. Antoci, private
communication):
\begin{equation}
r \to r'=r;~~ t\to t'=t; ~~ \phi \to \phi'=\phi; ~~ \theta \to \theta'=f(\theta)
\end{equation}
where $f(\theta)$ is a monotonic  function of $\theta$. It
can be seen that
while, Eq.(5)  remains unaltered under this transformation,
 Eq.(6) would not be valid unless one
fixes $f(\theta) = \theta$ uniquely. Physically, this unique choice of $\theta$
corresponds to measuring it from the axis of symmetry and ensuring that
$g_{\phi \phi} =0$ along the same and the azimuthal plane lies at $\theta =
90^o$. Whenever we adopt this convention of defining $\theta$ uniquely,
Eq.(6) would be applicable. On the other hand, if we do not fix $\theta$ uniquely,
as mentioned above, i.e, even if we allow the limited freedom of
$\theta \to \theta' = \theta + \theta_0$, Eq.(6) cannot be invoked.

This form of the metric, known as the {\em Standard Form}, 
simplifies the Einstein equations considerably and
is widely used for studies of rotating compact objects[9], stationary
axially symmetric rotating wormholes[10], and above all, the rotating black holes[1].

It may be emphasized that the general form of stationary axisymmetric
metric (in the Standard Form) widely used in the literature is[2,3,9,10]:
\begin{equation}
ds^2 = e^\lambda dr^2 - e^\nu dt^2 + e^\mu r^2 [ d\theta^2 + \sin^2\theta
(d\phi - \omega dt)^2]
\end{equation}
where $\lambda=\lambda(r, \theta)$, $\nu=\nu(r, \theta)$, $\mu =\mu (r, \theta)$.
And $\omega =\omega(r, \theta)$ is associated with the Relativistic
Frame Dragging effect.

Note that Eq.(6) is already incorporated in the foregoing form, i.e, the
physical conditions of asymptotic flatness  and measurement of $\theta$
w.r.t. the axis of symmetry are already imposed on the metric.

The Kerr metric, in the so-called Boyer and Lindquist[1] coordinate has
also this {\em Standard Form}:
\begin{eqnarray}
ds^2 & = & [r^2 +a^2 +{ 2mr\over \rho^2} a^2 \sin^2 \theta] sin^2\theta d\phi^2 
+ {\rho^2\over \Delta} dr^2 +\rho^2 d\theta^2  \nonumber\\
& & + {4amr\over \rho^2} \sin^2\theta d\phi dt  -(1- {2mr\over \rho^2}) dt^2 
\end{eqnarray}
Here $m$ is the mass of the Kerr BH, $a$ is the angular momentum per unit mass
and
\begin{equation}
\rho^2 = r^2 + a^2 \cos^2\theta; \qquad \Delta =r^2 -2mr +a^2
\end{equation}
It is easily seen that Eq.(9) is in the form of Eq.(7)
  where all cross terms except the one containing
$d \phi dt$ are $0$ and where

\begin{equation}
g_{\phi \phi} = [r^2 +a^2 +{ 2mr\over \rho^2} a^2 \sin^2 \theta] \sin^2\theta 
\end{equation}

and

\begin{equation}
g_{\theta \theta} = \rho^2
\end{equation}

Metric.(10) {\em too uses} $x^2=r$ and $x^3=\theta$, is also
 asymptotically flat and  measures $\theta$ uniquely 
w.r.t. the axis
of symmetry. 

For instance suppose we subject Eq.(12) to the coordinate transformation (8).
While the LHS $g_{\phi \phi}\to g'_{\phi \phi}= g_{\phi \phi}$ would remain unchanged
under this transformation, the RHS will not do so unless $\theta'=f(\theta)=
\theta$ is already fixed uniquely. And unless we measure $\theta$ in this
unique fashion, $g_{\phi \phi}$ may not vanish along the axis of symmetry.
It is also known that when one uses the Boyer-Lindquist coordinate, the azimuthal
plane lies at $\theta = 90^o$ which is also the case with the original
Kerr coordinates[11] (but the original Kerr metric is not in the standard form and Eq.6)
cannot be invoked there).
Thus Eq.(6) should be applicable to the
Boyer-Lindquist metric but not to the original Kerr metric[11]
(see a cross-check to this effect at the end of the paper).

Then combining Eqs. (6), (12) and (13), we obtain

\begin{equation}
[r^2 +a^2 +{ 2mr\over \rho^2} a^2 \sin^2 \theta]
\sin^2\theta  = \rho^2 \sin^2\theta
\end{equation}
In the foregoing Eq., by first cancelling $\sin^2\theta$ from both sides, 
then using Eq.(11) on the R.H.S., then transposing and finally using the
identity $\sin^2\theta +\cos^2\theta =1$, it follows that
\begin{equation}
a^2 \sin^2\theta ( 1+ 2mr/\rho^2) =0
\end{equation}
The Eq.(15)  tells that either
\begin{equation}
a=0;\qquad  m \ge 0
\end{equation}
or,
\begin{equation}
a\ge 0; \qquad 2mr =-\rho^2; \qquad m\le 0, \qquad if ~ r\ge 0
\end{equation}
These foregoing conditions show that there cannot be any rotating BH with
$m\ge 0$ and if it is assumed that the compact object is strictly
non-rotating, i.e, $a=0$, only then it is possible to have $m \ge 0$.
Hence if the astrophysical BH candidates having $m >0$ would at all be
BHs, must be strict Schwarzschild ones.  Further, probable occurrence of
Kerr BHs with negative masses, as apparently allowed by Eq.(17) can
actually be ruled out on the following grounds: Combining Eqs. (11) and
(17), one would have
\begin{equation}
r^2 + 2m r + a^2 \cos^2\theta =0
\end{equation}

or,
\begin{equation}
r = -m \pm \sqrt{m^2 - a^2 \cos^2 \theta}
\end{equation}
This is the equation of a 2-D surface and thus if there would be a negative mass Kerr BH,
the associated 4-D spacetime would collapse to a 3-D spacetime. Hence negative masses
(i.e, Eq.[17]) can be rejected in the present context even without invoking
any ``Positive Energy Theorem''.

Therefore, we find that although the Kerr solution appears very appealing, 
{\em in a strict sense},
it is not so  because it cannot describe the gravitational field
of any spinning object of finite mass, not even the supposed BHs.

 Newman and Janis[12]
derived the Kerr metric by starting from the Schwarzschild form 
in a method which is ``curious'' as 
admitted by themselves. Recall that the radial variable appearing
in the Schwarzschild metric $R$ is very much a real quantity and is in fact a 
scalar too. Newman and Janis first effected a coordinate transformation of the
form:
\begin{equation}
R \rightarrow r\qquad \theta^\prime =\theta_{Schwarzschild}
\end{equation}
where $r$ is {\em allowed to be a complex variable}. After this, they 
introduced another
transformation of the form
\begin{equation}
r^\prime = r + i a \cos\theta; 
\end{equation}
where $r^\prime$ is seen to be the radial coordinate of the Kerr metric.
In effect, Newman and Janis 
{\em pretended}  as if $r^\prime$ were a real variable. But a careful consideration
would convince that no physically allowed coordinate transformation can transform
a purely real variable into a complex one and thus $r$ must be a real variable under
physically admissible transformations.  Consequently 
 $r^\prime$ can truly be a real variable
iff $a=0$ which is our Eq. (16). In fact Newman and Janis admitted that

\begin{verbatim}
``there is no simple, clear reason for the series of operations 
performed on the tetrad to yield a new (different from Schwarzschild) 
solution..''.
\end{verbatim}

 On the other hand, Chandrasekhar[13]  derived the Kerr metric by starting
with a general axisymmetric stationary metric of the form (7) (but in
cylindrical coordinates), and one may wonder why such a general approach
too should eventually lead to apparent incongruities like Eqs. (16-17).
While deriving the Kerr metric, Chandrasekhar not only used the justified
condition that for an empty spacetime, energy momentum $=0$, but he also
{\em assumed beforehand} that the resultant vacuum spacetime must contain
an Event Horizon (EH), mathematically, a ``null surface'' spanned by two
Killing vectors corresponding to $\phi$ and $t$ symmetries. This is in
contrast to the derivation of the vacuum Schwarzschild metric where one
does not, beforehand, force the existence of any EH and where the EH
arises on its own.  This suggests that while a spherically symmetric
vacuum spacetime does allow the existence of an EH for $m\ge 0$ (as long
as we ignore associated physical problems) a stationary axisymmetric
vacuum spacetime does not do so  even when we consider only mathematical
symmetry arguments (as long as we insist for $m\ge 0$). If there would be
a Kerr BH with $m >0$, as noted by Carter[14], there would be very severe
violation of causality for the internal solutions because of the
occurrence of non-removable closed timelike curves in regions of finite
positive $r$.  Consequently, he noted that there would be a ``breakdown of
general relativity'' and ``the whole theory might have to be abandoned''.
But we see here that there is no ``breakdown of general relativity'' and
the apparent ``breakdown'' was either due to pretentious mathematical
``trics'' or because of undue assumptions. So what need to be
``abandoned'' are such aspects rather than ``the whole theory''.

Our result is probably in conformity with
Mach's principle in that for a purely empty spacetime, rotation cannot be meaningfully defined. It is also
in definite agreement with the recent detection of ultra-relativistic flow 
with bulk Lorentz factor ($\ge 10$)
from the Cir X-1, an object without an EH but with strong intrinsic magnetic 
field[14]. The latter aspect is
again in conformity
 with interpretation of recent observations of black hole candidates[5-8]: the
so-called Black Hole Candidates do possess intrinsic magnetic fields which the
astrophysical BHs cannot and hence the BH candidates are not BHs.
In practice, none of the astrophysical BH candidates or any other compact
object is expected to be strictly non-rotating and even if we ignore the
evidence for intrinsic magnetic moment in the BHCs, they can hardly be BHs
in view of Eq.(16).
On the other hand, they must be ultra compact objects with physical surface
and without any EH.

To summerize, we find that though the Kerr metric has fascinated both
relativists and astrophysicists for 40 years, as long as we are interested
in exact spacetime of finite mass spinning 
BHs or compact objects, Kerr metric is only of academic
interest. However, for compact objects spinning slowly, the far off
gravitational field might (or might not) be approximated by Kerr metric[2,
16].
Note that despite attempts made for almost 40 years, nobody has been able
to match the external/internal spacetime of any known physical body (other
than a Kerr BH) possessing pressure, temperature and positive mass-energy
density by the Kerr metric[2,16].

This is {\em not to tell that there cannot be spinning objects with valid physical
properties}; on the other hand, it only to remind that we have yet 
not been able
to  derive 
the {\em exact} form of the metric associated with any spinning object (except
Kerr BH). And it is a long  overdue task for the relativists to
derive the {\em exact} metric for a  spinning physical object (other than
Kerr BH). 
It is also not difficult to see why the Kerr metric is valid only
for Kerr BHs:

A spinning BH has no mass moment of order higher than $l=1$[2,16].
On the other hand, even if  a physical body with finite mass and extent is
perfectly spherical to start with,  it 
would develop deviation from spherical symmetry
once it starts spinning. And a spinning physical fluid with asphericity 
is likely to develop moments higher than the $l=1$[2].m For instance, one
can see the expression for quadrupole ($l=2$) deformity formula for an originally
spherical spinning fluid in Eq(3.60) of ref[2].

Our result that even a spinning BH cannot have finite mass can be {\bf
cross checked} in the following manner.  The Boyer-Lindquist metric(10)
can be most directly derived by using the Backlund transformation[17].
When one does so, the following relationship between $a$ and $m$ emerges
automatically:
\begin{equation}
a= m \sin\phi
\end{equation}
Since $a$ and $m$ are constants, but $\phi$ is a variable and $\sin \phi
\neq 0$ (in general), the foregoing equation
can be satisfied iff

\begin{equation}
a=m=0
\end{equation}
Thus, unequivocally, $a=0$ for Kerr BHs. And this result also confirms 
that the Boyer-Lindquist metric (10) incorporates the physical condition
 $f(\theta) =\theta$ as much as Eqs.(6) and (9) do. If one would go through
any article on BHs where Boyer-Lindquist coordinates are used one would find that
$\theta$ is uniquely measured from the axis of symmetry and the azimuthal
plane is marked by $\theta =90^o$, the prerequisite for Eq.(6).
 And since all astronomical
objects and physical objects (except few elementary particles) including
the BHCs have $m>0$, they are not Kerr BHs.  Thus neither the Quasars, nor
the micro-quasars, nor the Gamma Ray Bursts nor anything else is powered
by spinning Kerr BHs contrary to the present astrophysical paradigm.

However, it is possible that all such astrophysical objects are powered by
spinning BH Candidates which have physical surface and intrinsic magnetic
fields but no EH. In other words, all such powerful astrophyical central
engines might be powered, alongwith likely accretion power, by spinning
objects which are somewhat akin to (relativistic) pulsars. Such spinning
objects with physical surface and intrinsic magnetic fields, rather than
BHs from which nothing can escape (atleast classically), are most suitable
for understanding the origin of powerful collimated jets and radiation.
Such BH candidates, unlike cold Neutron Stars, will be ``hot'', i.e,
trapped radiation pressure will play an important role in supporting them,
The conventional mass upper limit of ``cold'' objects will be
irrelevant for them. 

Also, unlike strictly Neutron Stars, in strict  hydrostatic equilibrium,
these objects may be collapsing at an incredible slow rate and thus generate
radiation pressure/heat at their core by virtue of virial theorem (negative
specific heat). If any reader is desires to see Ref.[17] but may not have easy access
to it may request the author for a photocopy of the same.

\vskip 1cm
Acknowledgement: I thank Demos Kazanas (NASA, GSFC) and P.C. Vaidya (of
Vaidya Metric fame) for useful discussions and verifications. Although the
algebra leading to this result is simple and straightforward, it has
nevertheless been kindly verified by Zafar Ahmed, Subir Bhattacharya,
Nilay Bhat, S. Sahaynathan, Ravindra Kaul and Darryl Leiter.

\end{document}